\documentclass[usenatbib]{mn2e}
\usepackage{graphics,amsmath}
\begin{document}

\title[Optical and Near-IR Properties of UKIDSS-SDSS Quasars]
{The Optical and Near-Infrared Properties of 2837 Quasars \\
in the {\it UKIRT} Infrared Deep Sky Survey (UKIDSS) }

\author[Chiu et al.\ ]{Kuenley Chiu$^1$, Gordon T. Richards$^{2,3}$, Paul C. Hewett$^4$, Natasha Maddox$^4$ \\
%, Robert C. Nichol$^4$  \\
$^1$\,School of Physics, University of Exeter, Stocker Road, Exeter EX4 4QL, UK {\tt email: chiu@astro.ex.ac.uk}\\
$^2$\,Department of Physics, Drexel University, 3141 Chestnut Street, Philadelphia, PA 19104 \\
$^3$\,Department of Physics and Astronomy, Johns Hopkins University, 3400 North Charles Street, Baltimore, MD 21218\\
$^4$\,Institute of Astronomy, Madingley Road, Cambridge, CB3 0HA, UK
}

\date{Accepted for publication in MNRAS, December 19 2006}

\maketitle

\begin{abstract}
The UKIRT Infrared Deep Sky Survey (UKIDSS) is the first of a new
generation of hemispheric imaging projects to extend the work of the
Two Micron All Sky Survey (2MASS) by reaching three magnitudes deeper
in $YJHK$ imaging, to $K=18.2$ ($5\sigma$, Vega) over wide
fields. Better complementing existing optical surveys such as the
Sloan Digital Sky Survey (SDSS), the resulting public imaging catalogues
provide new photometry of rare object samples too faint to be reached
previously.  The first data release of UKIDSS has already surpassed
2MASS in terms of photons gathered, and using this new dataset we
examine the near-infrared properties of 2837 quasars found in the SDSS
and newly catalogued by the UKIDSS in $\sim$189\,deg$^2$.  The
matched quasars include the RA range 22$^h$ to 4$^h$ on the Southern
Equatorial Stripe (SDSS Stripe 82), an area of significant future followup possibilities
with deeper surveys and pointed observations.  The sample covers the
redshift and absolute magnitude ranges $0.08 \le z \le 5.03$ and
$-29.5 \le M_i \le -22.0$, and 98\,per cent of SDSS quasars have matching
UKIDSS data.  We discuss the photometry, astrometry, and various
colour properties of the quasars.  We also examine the effectiveness of
quasar/star separation using the near-infrared passbands. The combination
of SDSS $ugriz$ photometry with the $YJHK$ near-infrared photometry from
UKIDSS over large areas of sky has enormous potential for advancing our
understanding of the quasar population.
\end{abstract}

\section{Introduction}
\label{sec:INTRO}

In the several years since the SDSS (Sloan Digital Sky Survey) and
2MASS (Two Micron All Sky Survey) projects pioneered large uniform
digital datasets accessible to the astronomical community, another
generation of surveys has been spurred by improvements in imaging
technology, software pipelines, and the ability to carry out
sophisticated astronomical science on databases.  In particular, over
the next several years, two large consortia will be exploring deeply
into the near-infrared with surveys from 4-metre class telescopes.
Using UKIDSS \citep[UKIRT Infrared Deep Sky Survey,][]{lawrence06}
and VISTA \citep[Visible and Infrared Survey Telescope for Astronomy,][]{emerson04}, 
they aim to extend the limiting flux of the
2MASS $JHK$ bands by three magnitudes or more, as well as into new
parameter spaces using the novel $Z$ and $Y$ optical/near-infrared
filters \citep{hewett06}.

In part motivated by the search for objects such as the coolest brown
dwarfs and highest redshift quasars, deeper surveys in the
near-infrared provide valuable information for the practical selection
and identification of rare objects from parent catalogues of normal
stars and other contaminants.  Both high-redshift quasars and brown dwarfs are
marked by rising flux towards and beyond 1\,$\mu$m, but, until recently,
were rarely discovered in the near-infrared due to the shallow
limiting magnitude of existing surveys.  For normal Galactic
stars detected in SDSS, for example, 2MASS data has provided
sufficient depth for successful characterization, such as in
\citet{finlator00}, but brown dwarfs and high-redshift quasars are
typically much fainter than the 2MASS survey limits -- only $\sim$17\,per
cent of SDSS-discovered quasars are also detected in 2MASS, and these
are primarily at low redshifts.

With the first large data release by the UKIDSS, it is now possible to
study the near-infrared properties of a large sample of SDSS
spectroscopically identified quasars over an extended redshift range.
Further, with an eye towards future selection techniques, the addition
of deeper near-infrared photometry has potential for the
identification of new classes of rare quasars.  Just as importantly,
deeper near-infrared data facilitates the rejection of spurious
candidates in such surveys.  Therefore, combining data from the UKIDSS
with existing data from SDSS and other surveys may offer large gains
in observing efficiency.

In this work, we match spectroscopically identified
quasars from the SDSS DR3 with their newly released UKIDSS DR1
photometry.  We discuss the data matching procedures in the following
sections, examine the joint properties of the quasars compared with
normal stars, the photometric behaviour of the quasars versus
redshift, and their relation to synthetically generated model quasars.  As
in \citet{schneider05}, all absolute magnitude values in this paper
are based on the standard `concordance' cosmology of $\Omega_M=0.3$,
$\Omega_\Lambda=0.7$, $H_0=70$ km s$^{-1}$ Mpc$^{-1}$.

\begin{figure}
\hspace*{-0.3in}
\resizebox{0.53\textwidth}{!}{\includegraphics{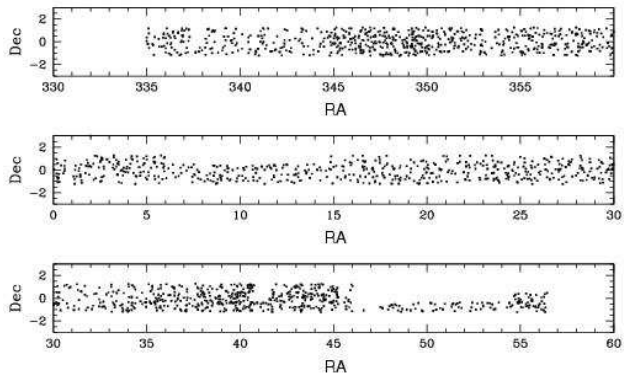}}
\vspace*{-1.5in}
\caption{Positions of SDSS DR3 quasars matched with UKIDSS catalog
sources are plotted in right ascension (RA) and declination
(Dec)(J2000).  The area plotted is the southern equatorial stripe (SDSS Stripe
82), a
representative subset of the total area matched in this work.  Some
obvious gaps in RA/Dec, around RA=50$^\circ$ for example, are due to
currently incomplete UKIDSS imaging, and the over-/under-density of
matched quasars versus RA ($\sim 30\%$) is due to the inhomogeneous
spectroscopic plate coverage of the input SDSS quasar catalogue.  ({\it Full-resolution figures will appear in published version, or may be downloaded in 
the meantime at http://www.astro.ex.ac.uk/people/chiu/chiu.figs.tar.gz })
\label{radecfig} }
\end{figure}

\begin{figure}
\hspace*{-0.1in}\resizebox{0.8\textwidth}{!}{\includegraphics{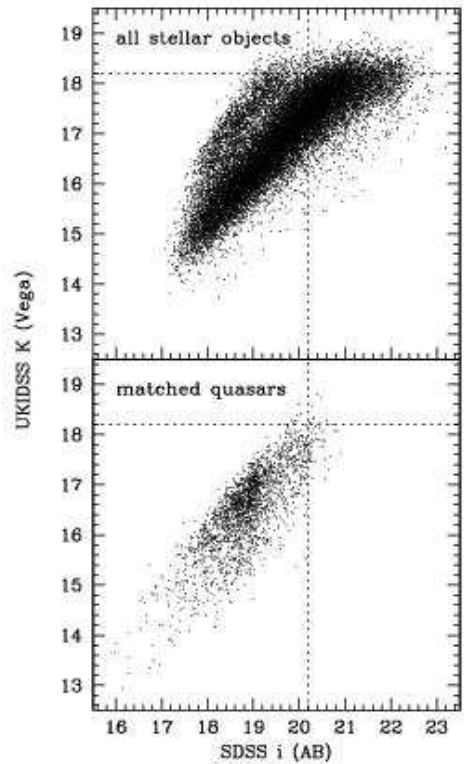}}
\vspace*{-1.3in}
\caption{SDSS-UKIDSS matched quasars (bottom panel) are shown compared
to a sample of raw unidentified point sources (top panel) that have
also been matched with UKIDSS.  Their apparent SDSS $i$ and UKIDSS $K$
magnitudes are plotted, illustrating the SDSS spectroscopic survey
limit ($i=20.2$, vertical dotted line) and UKIDSS photometric limit
($K=18.2$, horizontal dotted line).  ({\it Full-resolution figures will 
appear in published version, or may be downloaded in 
the meantime at http://www.astro.ex.ac.uk/people/chiu/chiu.figs.tar.gz })
\label{ikfig} }
\end{figure}

\begin{figure}
\resizebox{0.48\textwidth}{!}{\includegraphics{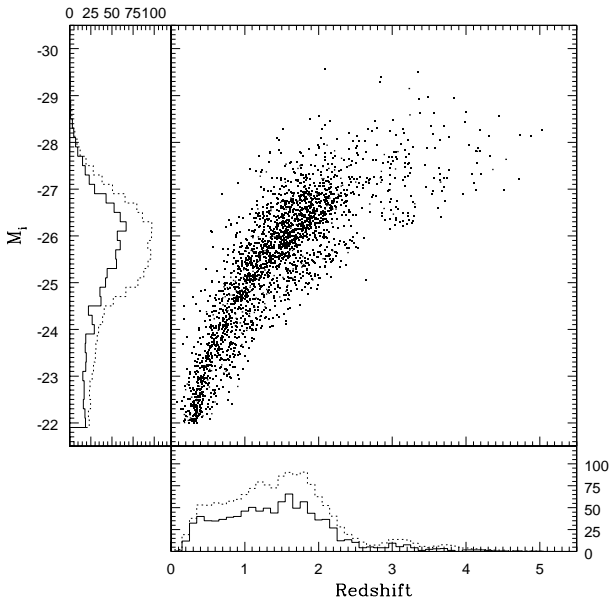}}
\caption{The redshift and absolute $M_i$ magnitude distribution of
UKIDSS-SDSS matched quasars is shown, and compared against the parent
SDSS DR3 quasar catalogue.  In the central panel are plotted the 2837
quasars from the successfully matched UKIDSS-SDSS subsample.  The
slight density edge perceptible within the sample results from the
magnitude limits of the SDSS main quasar spectroscopic survey
\citep{richards02}.  The side panels are histograms of the
distributions in redshift and absolute magnitude, with the parent DR3
quasar catalog shown as a dotted line, and matched quasars as a solid
line.  The parent sample histograms have been scaled by a factor of
0.4 for comparison.  \label{redshiftmagdistfig} }
\end{figure}

\section{Data Sources}
\label{sec:datasources}

\subsection{SDSS DR3 Quasar Catalog}
\label{sec:dr3}

The quasars used in this work are drawn from the SDSS DR3 Quasar
Catalog\footnote{http://www.sdss.org/dr5/products/value\_added/qsocat\_dr3.html},
assembled by \citet{schneider05}.  The DR3 quasar catalogue was
generated from spectroscopically confirmed quasars discovered and/or
reidentified by the SDSS over 3732\,deg$^2$, within the main SDSS
northern area as well as southern equatorial stripe (SDSS Stripe 82).  
While futher
updates to the SDSS quasar catalogue are ongoing, the DR3 version
represents the final product in the right ascension (RA) and
declination (Dec) regions used in this paper.  The SDSS automated
spectroscopic quasar program, whose selection criteria and strategy
have been discussed in \citet{richards02}, targets quasars at
redshifts up to $z=5.4$, and down to magnitudes $m_i<20.2$ (AB), using
several magnitude and colour cuts applied to the SDSS $ugriz$
photometry.

Following inspection of the individual spectra and compilation with
attributes such as radio, near-infrared, and X-ray emission
properties, Schneider et al. retained 46,420 objects, ranging in
redshifts from $z=0.08$ to $5.41$, and absolute magnitudes $M_i=-22$
to $-30.2$ (including some serendipitously discovered quasars down to
a spectroscopic survey limit of $i<21.0$).  Typically, the photometric
accuracy of the SDSS is $0.03$\,mag and systematic
astrometric errors are less than $0\farcs2$.  These photometric and
astrometric properties reflect the high quality of SDSS data. Further
details of the SDSS hardware and software can be found in:
\citet{york}, \citet{fukugita}, \citet{gunn98}, \citet{hogg},
\citet{lupton}, \citet{stoughton}, \citet{smith}, \citet{pier},
\citet{abz}.

\begin{figure}
\resizebox{0.48\textwidth}{!}{\includegraphics{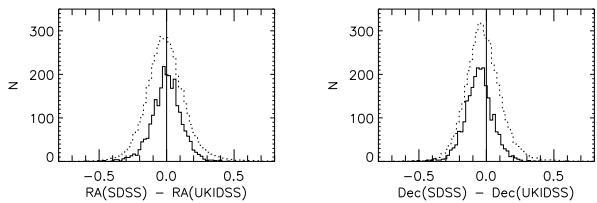}}
\vspace*{-1.2in}
\caption{Differences between SDSS and UKIDSS measured positions in
 arcseconds.  The dotted line shows results for a bright stellar reference
 population of 34,000 sources in the matched area.  Solid line shows
 the sample of 2837 matched quasars.  The modeled profiles have
 Gaussian FWHM of $0\farcs3$ for the quasar sample, and $0\farcs35$
 for the stellar reference sample.  The stellar population histograms
 have been scaled by a factor of 0.14 for comparison.
 }\label{posdiff}
\end{figure}

\subsection{UKIDSS}

UKIDSS represents the first wide and deep ground-based survey in the
near-infrared regime since 2MASS operations concluded in 2001.  The
UKIDSS\footnote{http://www.ukidss.org} is composed of several
sub-surveys targeting different parameter spaces of both Galactic and
extragalactic observations, planned in a typical `wedding cake'
strategy which balances wide-field versus deep coverage.  The five
sub-surveys include deep/narrow target fields such as the UDS (Ultra
Deep Survey), and wide/shallow fields such as the GPS (Galactic Plane
Survey).  Of relevance to this work is the UKIDSS Large Area Survey
(LAS), which aims to cover an area of 4,000 deg$^2$ matching the SDSS
over seven years.

 During the $\sim130$ nights per year in which the WFCAM instrument
 \citep{casali01} is mounted on UKIRT and devoted to UKIDSS, the
 project observes the sky under tolerable seeing conditions
 ($<1\farcs2$), and photometric or mildly non-photometric skies.  Each
 pointing of the telescope covers an effective area of $0.21$ deg$^2$,
 but due to the sparse-filled arrangement of the four 2K$\times$2K
 infrared detectors, four pointings are necessary to fill in a full
 sky patch of $\sim 0.77$\,deg$^2$.  Because of the large pixel scale
 of the detectors ($0\farcs4$/pixel) and the large field of view,
 survey progress is quick, covering this filled tile area in the four
 LAS $YJHK$ filters to a depth of $K=18.2$ ($5\sigma$, Vega) in
 approximately 20 minutes, with a total of 40\,s effective
 exposure time per band.  Survey progress under good conditions can
 thus produce approximately 25\,deg$^2$ of complete LAS imaging
 data per night. \citet{dye06}, \citet{lawrence06}, and
 \citet{warren06} provide comprehensive descriptions of the survey
 hardware, software strategy and data product characteristics.
 
Following observation, pipeline processing, and catalogue generation,
the typical limiting magnitude of the LAS is $YJHK =
[20.16,19.56,18.81,18.19]$ (Vega, $5\sigma$) or [20.79, 20.49, 20.19, 20.09]
(AB, 5$\sigma$), with photometry having
been bootstrap-calibrated to the 2MASS system.  These limits represent
an approximately three magnitude depth gain compared to the 2MASS
$JHK$ imagery.  Astrometric accuracy in the LAS is typically
$<0\farcs1$, and photometric accuracy is $\sim0.04$\,mag, measured in
the $J$-band.  Imaging in the four filters is taken in pairs, $YJ$ and $HK$,
for any particular field. The imaging catalogues present essentially an 
instantaneous single epoch measurement for each filter-pair but significant 
time intervals can exist between the two filter-pairs.

\section{Catalogue Matched Data}

\begin{figure*}
\resizebox{0.95\textwidth}{!}{\includegraphics{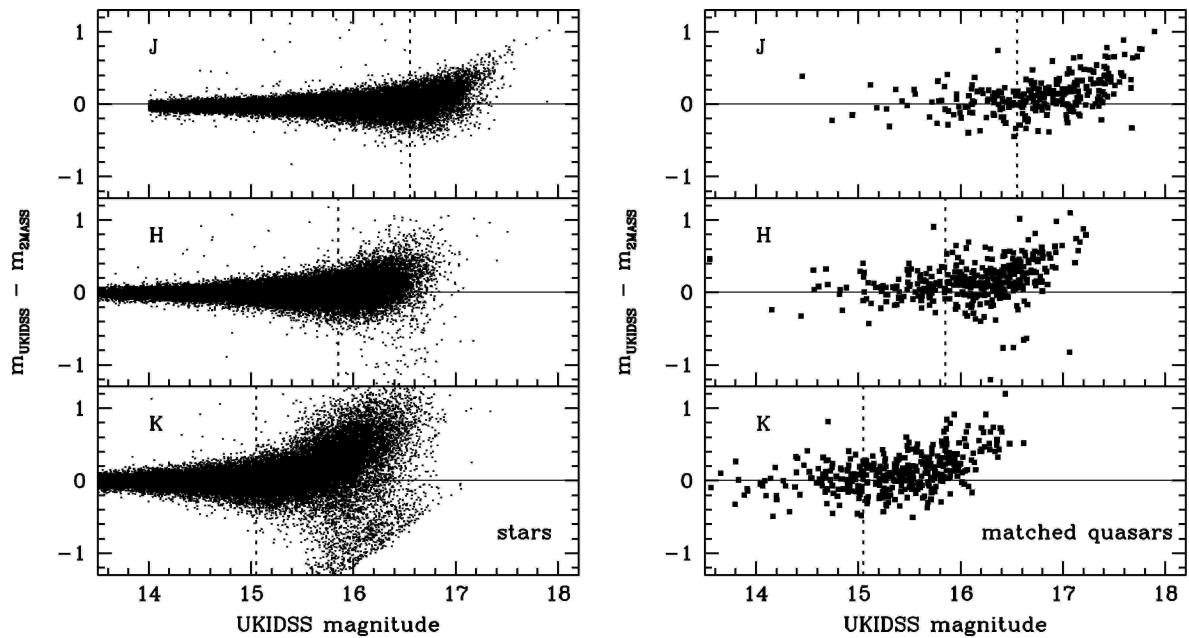}}
\vspace*{-0.01in}
\caption{Photometry differences between UKIDSS and 2MASS measurements.
 Left panel: for a sample of reliable stellar sources detected by both
 UKIDSS and 2MASS with $J>14$, the photometry differences between the
 two surveys are shown versus UKIDSS magnitude, in the three 2MASS
 bands, $JHK$.  Dotted vertical lines in both panels indicate the
 2MASS $5\sigma$ limiting point-source magnitude in each band ($JHK =
 16.55, 15.85, 15.05$, Vega).  Right panel: the photometry of 474 (out of
 2837 matched) SDSS DR3 quasars found in both UKIDSS and 2MASS
 catalogues is shown, illustrating the sample scatter and that most SDSS
 DR3 quasars will be found below the 2MASS catalogue limits.  ({\it Full-resolution figures will 
appear in published version, or may be downloaded in 
the meantime at http://www.astro.ex.ac.uk/people/chiu/chiu.figs.tar.gz })
 }\label{magsdifffig}
\end{figure*}

\subsection{Matching Procedure}

The UKIDSS DR1 release overlaps a subset of the SDSS northern and
southern areas and achieves a limiting magnitude nearer to the SDSS
for a typical quasar spectral energy distribution (SED) than did
2MASS.  At the same time, the photometric and astrometric performance
of the UKIDSS is similar to the SDSS.

The areal extent of the newly released UKIDSS imaging is much
smaller than the available SDSS DR3 coverage and we first trimmed the
Schneider et al. (2005) quasar catalogue to dimensions similar to
the UKIDSS DR1 imaging areas.  Among the total UKIDSS DR1 imaging
area released, a significant portion is located in the southern
equatorial stripe, which is also a special SDSS region where many
repeated scans have been taken over the lifetime of the survey.  In
the future, this area may become a target of deeper survey programs at
various wavelengths.  The SDSS and UKIDSS have a common southern
equatorial stripe area with an RA-range of approximately 22$^{\rm h}$
to 4$^{\rm h}$, and a declination-range of $\pm 1.265^\circ$, centered
on 0$^\circ$.

Extracting quasars from the DR3 catalogue without any imposed
constraints, the quasars were then positionally
cross-matched with the UKIDSS DR1 database, with a tolerance of
$1\farcs0$, matching only the nearest object to the queried
position. A total of 2837 SDSS quasars were succesfully matched to the
UKIDSS DR1 database.  

\subsection{Matched sample properties}

\subsubsection{General properties} A portion of the southern equatorial
RA/Dec distribution of the resulting sample of SDSS-UKIDSS
quasars is shown in Figure \ref{radecfig}.  Other sky-areas contributing 
to the SDSS-UKIDSS sample have more complicated geometries \citep[described in][]{warren06}, 
but show similar matched object densities (within
$\sim30$\,per cent due to SDSS spectroscopic plate coverage
over-/under-densities). Within the RA-range 1$^{\rm h}$ to
$2^{\rm h}40^{\rm m}$, of 574 input quasars in the southern equatorial stripe,
(corresponding to a sky density of $9 ~{\rm deg}^{-2}$), 567 (98.7\,per cent)
have a corresponding UKIDSS source with at least one detection in
the $YJHK$-bands. We employ the results of all the matches to the UKIDSS
DR1 catalogue, irrespective of the number of detections among the four
passbands. Figure \ref{ikfig} illustrates the relationship of
typical matched quasars to the parent sample of all matched point
sources in the joint SDSS-UKIDSS dataset, showing the spectroscopic
and photometric sample limits of the catalogues from which the quasars
are drawn.  Ongoing projects, for example employing the
novel AAOmega multifibre spectrograph at AAT \citep{saunders04}, aim
to extend the SDSS-selected quasar sample to magnitudes below these
limits, and improve constraints on the faint and distant quasar
luminosity function.

Figure \ref{redshiftmagdistfig} illustrates the redshift and absolute
magnitude dependence of the full sample of SDSS-UKIDSS quasars, which
show no strong bias from the parent unmatched sample of SDSS quasars.
Further, in UKIDSS areas which are imaged in all four UKIDSS $YJHK$
bands, we examined the frequency of quasar detection versus band and
found no strong dependence -- when a quasar is detected, it can
generally be detected in all four bands.  Specifically, for example in
the above representative area, out of 330 quasars found in the smaller
4-band imaged area, 329 are detected in the $Y$-band, 328 in the
$J$-band, 321 in the $H$-band and 325 in the $K$-band, while 319 have
detections in all $YJHK$ bands.

\subsubsection{Morphology} In the DR3 quasar catalogue, quasars at high
redshifts are generally unresolved by the SDSS imaging pipeline
\citep{chiu05}.  The SDSS-UKIDSS sample shows a similar behaviour in
the UKIDSS imaging bands, though with a greater fraction of resolved
quasars detected at low redshift -- 222 of 278 quasars with redshifts
$z<0.5$ are classified as extended in the UKIDSS {\tt mergedClass}
parameter.  Because the {\tt mergedClass} parameter combines weighted
measurements of the measured source profile from each of the available
bands, contribution from low redshift quasar host galaxies,
particularly in the $K$ band, may influence this trend (see Maddox \&
Hewett 2006).  However, by $z>1.0$, only $\sim$12\,per cent of quasars
are resolved in any UKIDSS band, falling to 7\,per cent and 5\,per
cent by $z>2.0$ and $z>3.0$ respectively.  In the present sample
(without regard to redshift), the majority of the quasars are
classified as point sources with good point spread function (PSF) fits
(UKIDSS database ${\tt mergedClass} =-1$, $N=1675/2837$), many are
classified as extended (${\tt mergedClass} =+1$, $N=717/2837$), and a
smaller fraction possess poor PSF-fits (${\tt mergedClass}
=-2$, $N=435/2837$).  Ten objects appear as uncertain
extended-source fits (${\tt mergedClass}=-3$), or are
classified as noise.

\begin{figure}
\hspace*{0.3in}
\resizebox{0.4\textwidth}{!}{\includegraphics{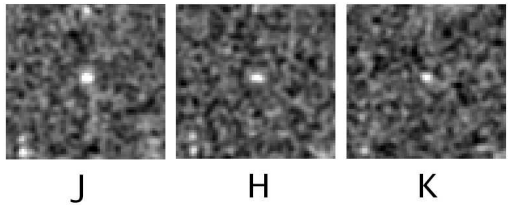}}
\caption{UKIDSS-detected but 2MASS-undetected quasars can be recovered
 in composite from 2MASS imaging.  60 SDSS-UKIDSS quasars with $J_{\rm
 UKIDSS}>18$ were selected and their 2MASS atlas images in $JHK$
 retrived.  Coaddition of the 2MASS-images yields the composite
 source above.  Each image is $1'$ square.  }\label{JHKstackfig}
\end{figure}

\subsubsection{Astrometry} ~~ As cosmologically distant point sources,
quasars should provide a fixed reference frame on the sky.  Therefore,
comparisons of the astrometry between different surveys of these same
objects can yield a measure of the relative astrometric errors.  Using
the sample of SDSS-UKIDSS quasars then, the positional offsets between
the surveys was determined.  The resulting distribution of relative
separations is shown in Figure \ref{posdiff}, along with a large
reference sample of reliably measured normal Galactic stars
($N\sim30,000$).  While a maximum offset of $0\farcs5$ can be seen for
a few quasars (and up to $0\farcs6$ for the stellar population), the
bulk of the sample is well modelled by a Gaussian distribution with a
full-width at half-maximum (FWHM) of $0\farcs3$.  In comparison, the
stellar population displays a somewhat wider distribution of
astrometric residuals, with a FWHM of $\sim 0\farcs35 $.  These
parameters are consistent with the SDSS/2MASS astrometric comparison
in \citet{finlator00}. We also find a small systematic absolute offset
in both RA and Dec of $\sim 0\farcs05$ between the SDSS and UKIDSS
positions.

\subsubsection{2MASS magnitudes} ~~ In addition to providing
SDSS-measured properties, the DR3 quasar catalogue was matched with
external datasets, such as FIRST and ROSAT detections, Galactic
extinction measurements and 2MASS photometry.  Using this information,
we examined the differences in quasar photometry between the 2MASS and
UKIDSS measurements.  For the SDSS-UKIDSS quasars, we extracted any
available 2MASS magnitudes in the $JHK$ bands.  As can be inferred
from the resulting sample in Figure \ref{magsdifffig}, many
SDSS-UKIDSS quasars fall well below the nominal 2MASS $5\sigma$
limiting magnitudes ($JHK = 16.55, 15.85, 15.05$, Vega).  However, a small fraction are still measured
in the 2MASS catalogues, down to approximately
$J\sim17.8$, $H\sim17.2$, and $K\sim16.5$.  In our sample, 474
of 2837 SDSS/UKIDSS matched quasars (17\,per cent) were sufficiently
bright to be detected in the 2MASS catalogue in any band.  Comparing
UKIDSS and 2MASS magnitudes in this range then, we find a scatter of
$\sigma = 0.3$\,mag between the two quasar measurements (above the
$5\sigma$ 2MASS limits), and similar to the envelope of scatter in
reference point sources at these faint magnitudes, which for example
ranges from $\sigma=0.1$\,mag at the bright end to $\sigma=0.4$\,mag
at the $5\sigma$ 2MASS detection limits (Figure \ref{magsdifffig},
left panel).  While quasar variability may account for a small portion
of the photometric differences, on the order of $\Delta m \sim
0.15$\,mag, as \citet{vanden04} have discussed previously, here the
photometric errors should dominate any such variability contribution.

Although many quasars are individually not detected in the 2MASS
imaging, the UKIDSS photometry now allows us to demonstrate the latent
detection ability of 2MASS.  Selecting 60 SDSS-UKIDSS quasars at
$J>18$ (which cannot be detected in 2MASS), we extracted the 2MASS
$JHK$ atlas images around each SDSS-UKIDSS quasar position.  The 2MASS
atlas images were then stacked, yielding the composite source shown in
Figure \ref{JHKstackfig}.  In areas of sky that will not be covered by
UKIDSS, or other future surveys, this technique may allow 2MASS to
provide useful information on populations of faint sources \citep[a technique 
recently demonstrated by][for example]{white06}.
 
\begin{figure}
\resizebox{0.48\textwidth}{!}{\includegraphics{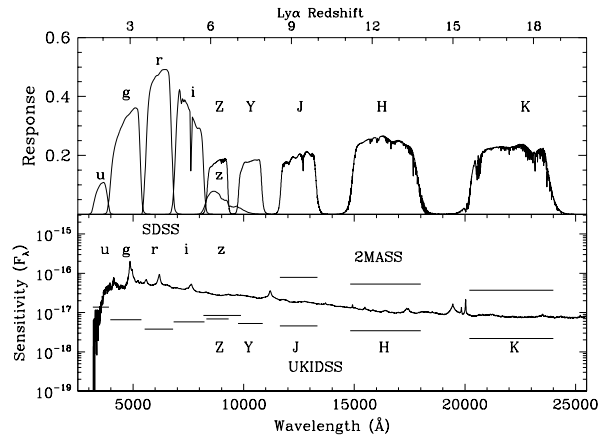}}
\vspace*{-2.2in}
\caption{Top panel: throughputs of the SDSS and UKIDSS imaging systems.
Note the novel UKIDSS $Z$ and $Y$ passbands, which have
sharper cutoffs compared to SDSS $z$, and fill the wavelength
gap between the SDSS and near-infrared $JHK$ bands.  For reference,
the top axis scale indicates the wavelength position of quasar
Ly$\alpha$ emission versus redshift.  Bottom panel: horizontal lines
indicate the resulting $5\sigma$ $F_\lambda$ sensitivities
(erg/s/cm$^2$/\AA) and wavelength coverages of the SDSS and UKIDSS
survey bands, in addition to the shallower 2MASS in $JHK$.  A
composite quasar spectrum shifted to redshift $z=3$ and normalized to
magnitude $i_{\rm AB}=19.1$ is shown, demonstrating near-infrared
detection in UKIDSS but not 2MASS.}\label{filtersfig}
\end{figure}

\section{Quasar colours} ~~ Using the SDSS-UKIDSS quasars, we now 
examine the optical and near-infrared colour properties of the sample, their 
relation to stellar properties in the same parameter space, and consider the 
utility of quasar-star separation.\label{magnitudescaveat}
An important preliminary note here is that in all calculations, as
well as resulting figures and tables in this work, we use the
magnitude systems returned automatically and without modification by
the SDSS and UKIDSS survey databases.  In the case of the SDSS, these
are asinh magnitudes on the AB system (specifically {\tt psfMags}),
while for the UKIDSS, these are Vega magnitudes calibrated to 2MASS
({\tt YJHKapermag3} for point sources).  We adopt this approach for
the ease of observers querying the joint
dataset\footnote{http://surveys.roe.ac.uk/wsa}, so that AB-to-Vega
magnitude and colour transformations (which are sometimes accidentally
neglected) are not necessary for immediate interpretation of retrieved
data in conjunction with results shown here.  Also, slight variations
in the choice of conversion factors (depending on the spectral
template used) will therefore not affect the results discussed.  For
subsequent analysis, however, a single more consistent physical AB
magnitude scheme may be preferable, using AB-Vega transformations such
as those tabulated in \citet{hewett06}.

\begin{figure}
\hspace*{-.55in}
\resizebox{0.57\textwidth}{!}{\includegraphics{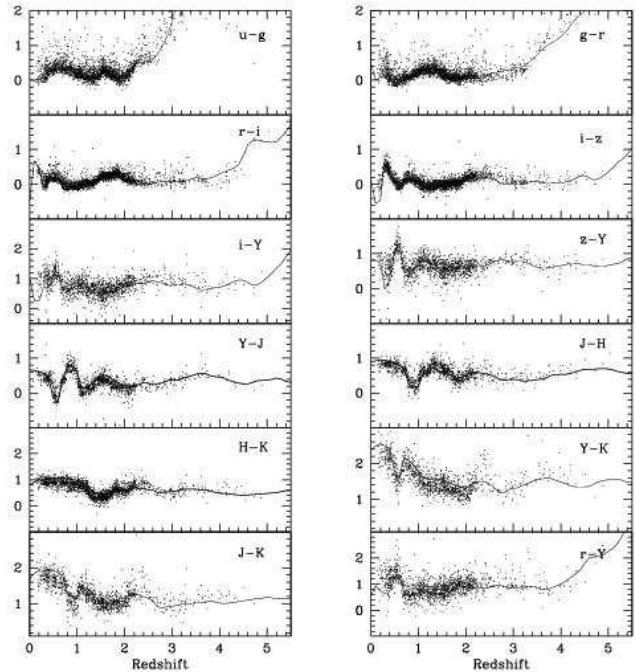}}
\vspace*{-0.3in}
\caption{ Colours of 2837 SDSS-UKIDSS quasars versus
redshift, in $ugrizYJHK$ combinations of interest.  The panels
generally progress from optical to near-infrared colours.  The colours
are calculated using the natural magnitude systems of each survey,
without conversion between AB or Vega (see \S\ref{magnitudescaveat}),
i.e. SDSS magnitudes are AB (asinh) magnitudes while UKIDSS
magnitudes are Vega-based.  The colour-redshift tracks of a composite
quasar are overlaid (solid line), generated from a modified version
of the model quasar SED descibed in \citet{maddox06}.  ({\it Full-resolution figures will 
appear in published version, or may be downloaded in 
the meantime at http://www.astro.ex.ac.uk/people/chiu/chiu.figs.tar.gz })} \label{coloursvszfig}
\end{figure}

\begin{figure}
\hspace*{-0.4in}
\resizebox{0.56\textwidth}{!}{\includegraphics{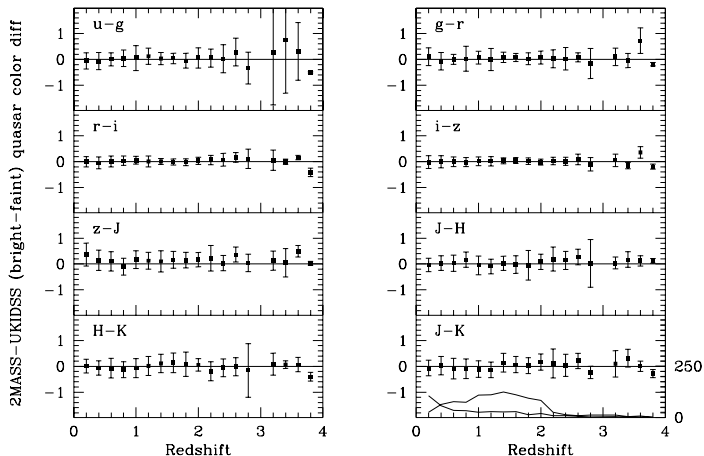}}
\vspace*{-1.3in}
\caption{The colour differences between 2MASS-detected (`bright') and
UKIDSS-detected (`faint', 2MASS-undetected) quasars are plotted,
versus redshift in bins of size $\Delta z=0.2$. 2MASS-detected quasars
are found up to $z=3.8$, providing the upper redshift limit to the
comparison sample.  Errors are calculated from the quadrature addition
of the standard deviations of the colour distribution in each 2MASS and
UKIDSS bin.  In the $J-K$ panel, two histograms (faint lines) show the
relative distribution of 2MASS and UKIDSS quasars contributing to the
comparison samples.  Typical numbers of quasars in the 2MASS sample
bins are $N=50$ at $z=0$, decreasing to $N=5$ at $z=3.8$, while
UKIDSS-only detected quasars peak around at $z=1.4$ with $N=130$.
Bins with insufficient numbers for comparison are omitted, such as at
$z=3.0$.}
\label{coloursdiffs}
\end{figure}

\subsection{The UKIDSS filter system}
The filter systems of the SDSS and UKIDSS are shown in Figure
\ref{filtersfig}.  In particular, while the SDSS $ugriz$ and MKO-based
$JHK$ filters are well-known and commonly used, the addition of the
novel $Z$ and $Y$ filters of the UKIDSS are a relatively recent
development and noteworthy.  Proposed by \citet{warren02}, and
discussed extensively in \citet{hewett06}, the UKIDSS $Z$ and $Y$
filters improved upon the existing broadband coverage between the optical
and near-infrared in several respects.

The $Z$ and $Y$ bands are defined by interference filters with sharp
wavelength cutoffs and, coupled with the high sensitivity of modern infrared
detectors, remedy some of the deficiencies in current imaging around
$\sim 1 \mu$m.  First, the UKIDSS $Z$ filter is much more sharply
defined than the SDSS $z$, because the SDSS $z$ filter itself in fact
has a relatively flat transmission of $\sim 0.4$ which continues out
beyond $1\mu$m \citep[illustrated by][]{fukugita}.  The overall 
$z$-band redward cutoff (i.e. total system throughput) then relies
on the declining CCD silicon sensitivity to terminate the band, and
thus results in a very extended filter profile. The CCD-detector
sensitivity envelope, beginning with a minimum in the ultraviolet, rising 
to a maximum around the $r$-band, and terminating with a minimum at
$1 \mu$m, largely accounts for the highly peaked shape of the SDSS
throughput.  This is in contrast to the relatively flat response of
the UKIDSS system, owing to the roughly constant sensitivity of the
HAWAII-2 HgCdTe detectors throughout the near-infrared.

Continuing redward, the $Y$-band (used by the UKIDSS LAS) provides a
new region of sensitivity in the $\Delta \lambda\sim 1500$ \AA ~
wavelength gap between the $z$(or $Z$)-band and the MKO $J$
band.  Among other motivations, the $Y$ filter was specifically
designed with the selection of very cool brown dwarfs and
high-redshift quasars in mind.  The increasingly red flux of these
objects versus temperature (for brown dwarfs) or redshift (for
quasars) has proven highly useful in rare object searches in the last
few years \citep{chiu06,fan06,chiu05}, but has reached a barrier as
current surveys, such as the SDSS, no longer provide sufficient
sensitivity in the $1 \mu$m range.  The high throughput of the UKIDSS
system at these wavelengths, combined with sharper band profiles,
allow the more precise discrimination of potential quasar candidate
redshifts and brown dwarf types, while avoiding unnecessary sky
background contamination.

The resulting $F_\lambda$ sensitivity of the UKIDSS imaging is shown in the
bottom panel of Figure \ref{filtersfig}, and compared with the SDSS as well as
2MASS sensitivities in the $JHK$ bands.  A composite quasar spectrum at $z=3$ is
also shown, illustrating the important capability of UKIDSS to detect the
typical quasar near-infrared continuum, versus non-detection in 2MASS.

\subsection{Quasar colours vs. redshift}

In Figure \ref{coloursvszfig}, we plot the 2837 SDSS-UKIDSS quasars, in various
colours versus SDSS spectroscopic redshift.  In each colour, only those quasars
with detected flux in both of the two contributing bands are plotted, i.e.
objects with non-detections in the necessary filters (indicated as such as by
${\tt mag}=-9999.99 $ in the SDSS or UKIDSS catalogues) are omitted.

As has been discussed in \citet{richards02}, \citet{chiu05}, \citet{maddox06},
and \citet{richards06a}, the behaviour of quasar colours is suprisingly
well-modeled across extended redshift and luminosity ranges by a single
composite SED.  The passage of the strong emission lines in and out of the
various filters controls the small-scale fluctuation of the quasar colours
versus redshift, while the extended underlying continuum provides the
distinguishing characteristic separating quasar colours from those of stars.
Importantly, for certain applications, such as the $z>6$ $i$-dropout selection
used by \citet{fan06}, absorption by the intervening intergalactic medium
provides a strong spectral discontinuity allowing the identification of quasars.
This effect can be seen for example in the rapidly rising $r-i$ colour at redshifts $z>4$,
as more quasar flux in the $r$ band is absorbed by the intergalactic medium with
increasing redshift.

The observed quasar sample beyond $z\sim2.5$ is affected by the declining space
density of the quasar population and the small volumes probed.  As a result, the
flux-limited sample becomes quite sparse (Figure \ref{coloursvszfig}).  In order
to extend the useful predicted colour-redshift values further than $z\sim2.5$,
we generated synthetic photometry and colours of a composite quasar spectrum to
predict the median redshift-colour evolution track.  The technique has been
described previously \citep[`quasar cloning' -- ][]{maddox06, richards06a, hewett06, chiu05}, and we follow a similar approach here.  The modified composite quasar
based on the spectrum of Maddox \& Hewett was employed as the model SED, updated
to include a base continuum contribution, which dominates at $\lambda <
10000$\,\AA, consisting of a dual power-law form \begin{equation} \alpha =
\begin{cases} -0.5 &\lambda <2600 {\rm \AA}\\ -0.2 &\lambda >2600 {\rm \AA},
\end{cases} \end{equation} where $\alpha$ specifies the quasar continuum slope,
$f_\nu \propto \nu^{\alpha}$.  A blackbody component, with T$=1850$\,K,
dominates at $\lambda > 10000$\,\AA, resulting in an overall spectral shape
similar to the composite spectrum of Glikman et al.  (2006).  For each redshift
bin in steps of $\Delta z=0.1$, the model SED was redshifted and the average
effect of intervening Lyman-$\alpha$ absorption added.  The resulting quasar
spectra were used to generate magnitudes and colours, again with SDSS magnitudes
on the AB system, and UKIDSS magnitudes on the Vega system, as would naturally
be found in the respective survey databases.  These are overlaid in Figure
\ref{coloursvszfig}.  Similar colour-redshift loci, based directly on the
various composite quasar spectra available \citep[e.g.][]{francis01,
brotherton01,vanden01}, are frequently employed for a
variety of purposes.  It should be stressed that the significant contribution of
host galaxy light in the low-luminosity quasars that typically dominate the
composite spectra at wavelengths $>5000\,$\AA \ leads to predictions that are
far too red in the optical-near-infrared colours, compared to the observations,
at redshifts $z > 2$.

In addition, given the large magnitude range covered by the SDSS-UKIDSS quasars,
we investigated whether the observed distribution of quasar colours showed
dependence on luminosity.  This was accomplished by selecting two subsets of
quasars, those bright enough to be detected in 2MASS, and the remaining ones
detected only in UKIDSS.  The colour differences between the two samples were
calculated in redshift bins of size $\Delta z=0.2$ and are shown in Figure
\ref{coloursdiffs}.  No significant difference is seen between the two samples.

\begin{figure*}
\resizebox{0.95\textwidth}{!}{\includegraphics{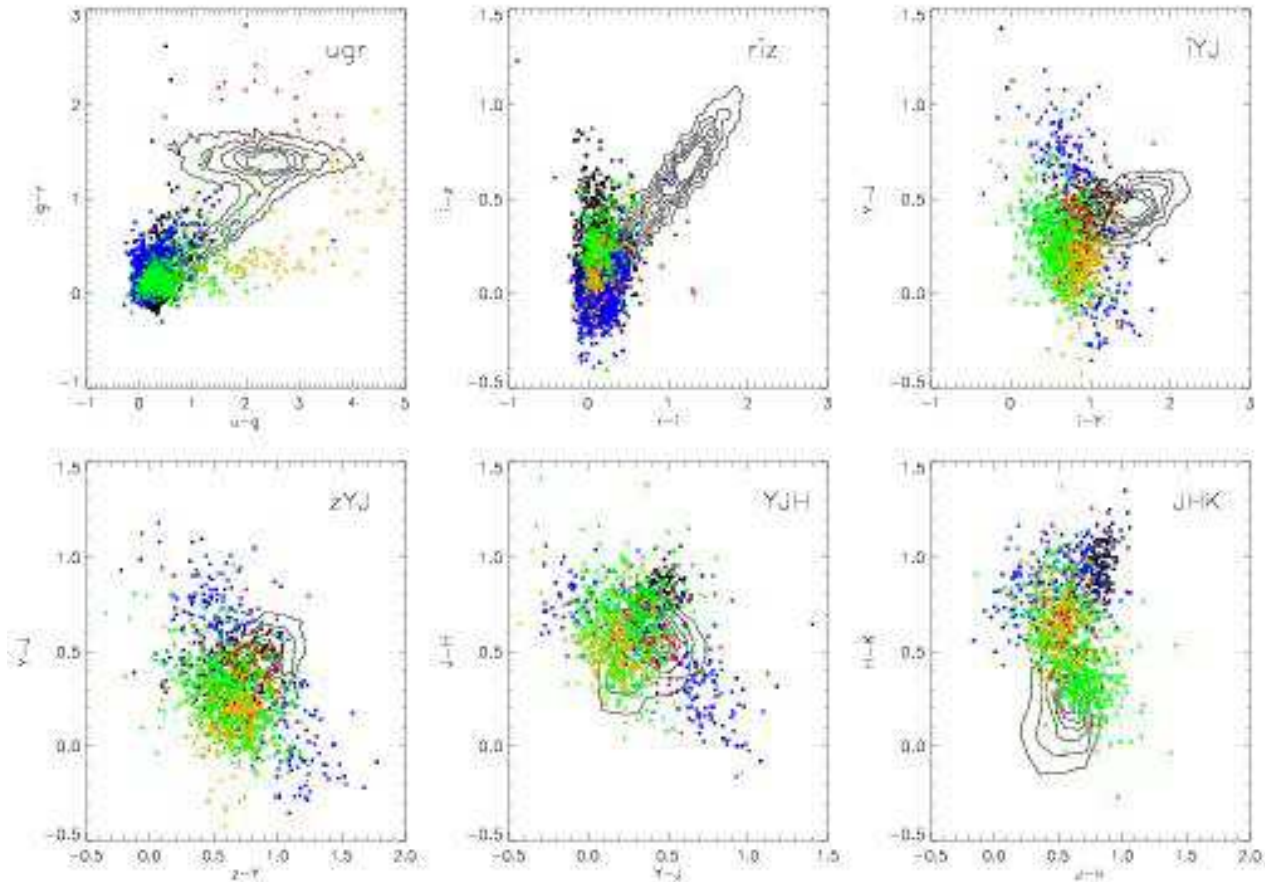}}
\caption{SDSS-UKIDSS quasars (small dots) in colour-colour diagrams
($ugrizYJHK$) with the density of normal stars shown as contours.  Quasar-star
separation is highly effective in colours involving the bluest bands, such as
$u-g$, but then decreases in utility in the optical.  With near-infrared and
particularly $K$-band imaging, quasar-star separation again can provide a
powerful selection method.  {\it Note}:  colours here are calculated from
natural magnitude systems of each survey, without conversion between AB or Vega
-- {\it i.e.}  SDSS magnitudes are AB (asinh) magnitudes while UKIDSS magnitudes
are Vega based  (see \S\ref{magnitudescaveat}).  Quasar symbol colours are encoded as follows -- black:
$z<0.5$, blue:  $0.5<z<1.0$, green:  $1.0<z<2.0$, yellow:  $2.0<z<3.0$, red:
$3.0<z<6.0$)  ({\it Full-resolution figures will 
appear in published version, or may be downloaded in 
the meantime at http://www.astro.ex.ac.uk/people/chiu/chiu.figs.tar.gz })\label{colourcolourfig}}
\end{figure*}

\begin{figure} \resizebox{0.48\textwidth}{!}{\includegraphics{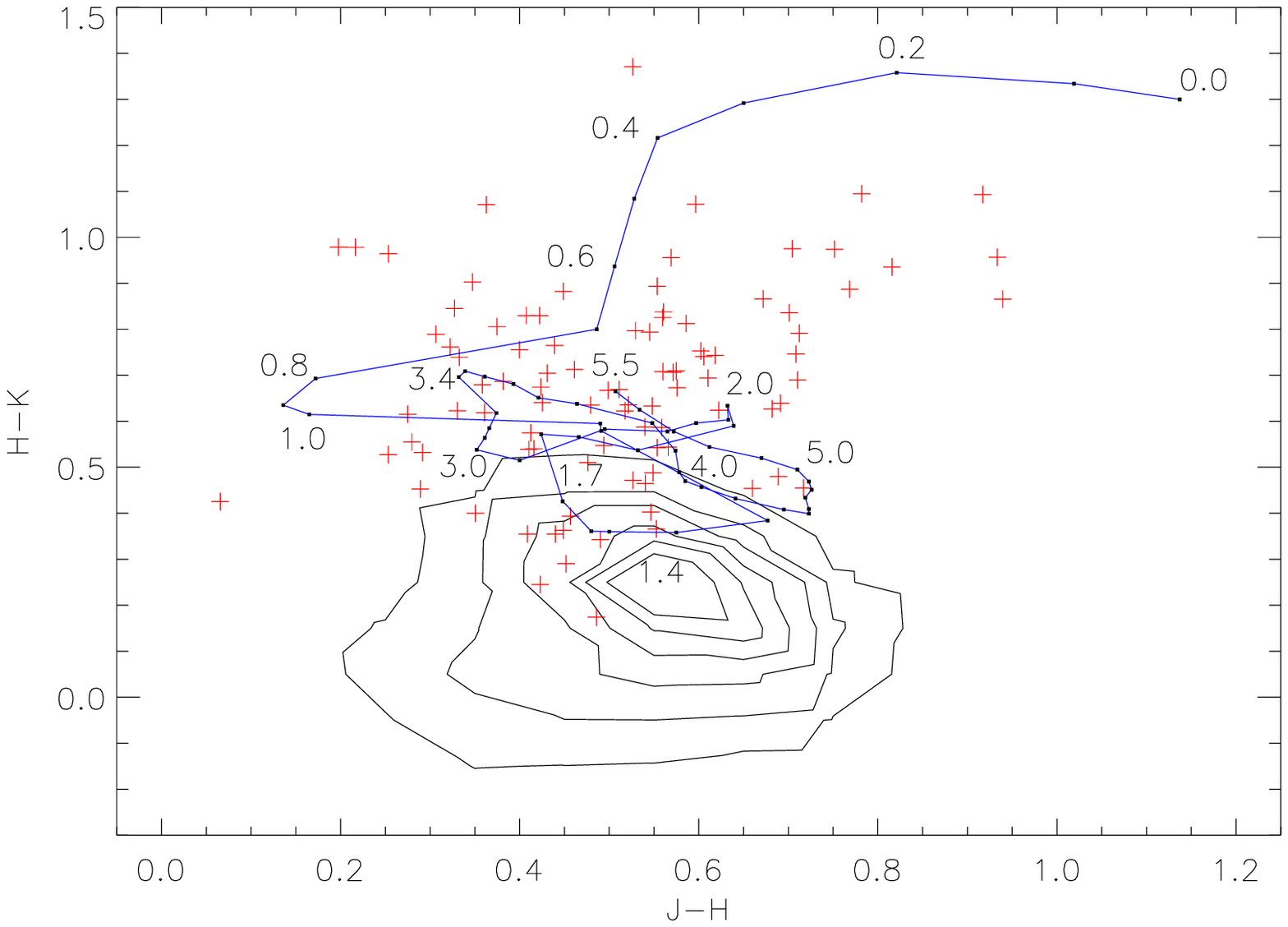}}
\vspace*{-0.1in} 
\caption{ A detailed view of the $(J-H)$, $(H-K)$ distribution of stellar
sources (contour lines) and SDSS-UKIDSS quasars at $2.2<z<3.3$ (red $+$
symbols), illustrating the useful colour separation of populations with $K$-band
information.  A model quasar track (discussed in text, blue solid line) is shown
from $z=0.0$ to $z=5.5$ with $\Delta z=0.1$ interval ticks and numeric labels.
The quasar track overlap with stellar sources is severe at $z\sim1.4$, but
reaches large separation from $z=2.0$ to $z=3.5$.}  \label{jhkfig}
\end{figure}

\subsection{Quasar colours versus normal stars}

To compare the quasar colours with those of normal stars, we extracted a stellar
sample from the UKIDSS/SDSS database.  This was accomplished by querying the
UKIDSS {\tt reliableLasPointSource} table, which returned relatively bright
objects displaying good photometric errors ($\sigma<0.1$ mag), unresolved
shape/flux profiles, and in uncrowded areas.  Then, for a few selected
two-colour spaces of interest, the stellar and quasar samples are plotted in
Figure \ref{colourcolourfig}.

Photometric searches for quasars generally compete against heavy contamination
by stellar objects with similar colours, due to the intrinsic rarity of quasars.
The relative distributions of quasars versus stars in a chosen colour
combination will generally control the search efficiency, i.e.  how many
contaminants must be observed to return one object from the desired sample.  The
progression of the two-colour plots in Figure \ref{colourcolourfig} illustrates
the varying success of distinguishing quasars from normal stars using different
colours.  For colours involving the ultraviolet, such as in $(u-g)$, $(g-r)$,
the strong short-wavelength quasar continuum flux is well known to provide a
separation from stars via the ultraviolet-excess (UVX) technique.  Relatively
clear separation of the two populations is observed here and provides a highly
successful selection method for low redshift quasars.

In the optical bands, adjacent filter combinations provide little discrimination
between the quasar and stellar populations.  However, with the addition of
near-infrared data, and particularly $K$-band imaging, the extended quasar
continua again results in an effective differentiation between quasars and
stars.  This attribute, exploited by the `KX' method \citep{warren00}, can
provide a useful selection at high redshift, $z \ge 2.5$, where the UVX
technique is unsuccessful due to the effect of absorption by the IGM in the
bluest passbands.

Looking farther into the infrared, an additional area of significant interest in
the future may be the combination of deep near-infrared surveys such as UKIDSS
with mid-infrared legacy imaging from the {\it Spitzer Space Telescope}.  As has
been demonstrated by \citet{stern05}, \citet{hatz05}, and \citet{richards06b},
quasar-star separation is highly successful in the mid-infrared bands, and also
benefits from high sensitivity due to the supressed background ($15 \mu$Jy in
$90$\,s at $3.6 \mu$m).  In areas with good optical, near-infrared, and proposed
mid-infrared imaging coverage from {\it Spitzer}, such as the southern
equatorial stripe, the combination of the resulting catalogues has the potential
to reach new distant and faint quasar populations while providing nearly
continuous bandpass coverage as emission lines and continuum flux shift with
redshift.

\section{Quasar colour selection strategies}

The utility of quasars in probing the distant universe has been demonstrated to
great effect in the past few years, with these high redshift objects
illuminating not only their own properties, but also the physical state of the
IGM and the distant universe.  However, the discovery of quasars in certain
redshift intervals is, in practice, a laborious enterprise.  In the absence of
information from the wavelength extremes, identifying quasars by special
properties such as radio or X-ray flux, discovery of quasars at redshifts near
$z \sim 2.5$, $z\sim 5.6$, and $z>6$ remains a significant challenge.  Observers
fight the combination of the intrinsically declining numbers of the target
population and the increasing numbers of contaminating Galactic stars.  In
previous work we have outlined some of the selection techniques and numbers of
quasars to be expected in the SDSS and UKIDSS, and have found that the
UKIDSS-SDSS joint dataset is capable of selecting quasars at $z>6.5$
\citep{hewett06,chiu05,warren02}.  However, at the magnitude limits of the
UKIDSS ($m_{AB}\sim 20-21$), the $N\sim90$ expected quasars at $z>6.5$ on the
whole sky clearly places this parameter space in the category of rare,
high-value, but resource-intensive targets.

\begin{figure} \resizebox{0.48\textwidth}{!}{\includegraphics{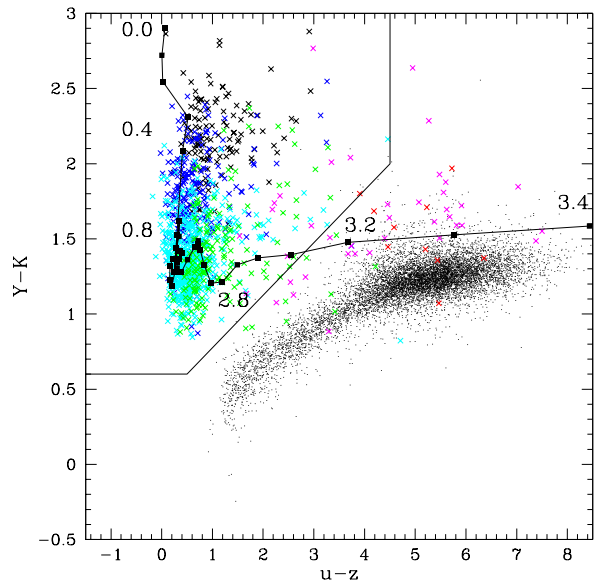}}
\vspace*{-0.2in} 
\caption{ An example of a successful low-redshift quasar selection using colour
separation in $(u-z)$,$(Y-K)$ space is illustrated.  Normal stars are plotted as
small dots, while SDSS-UKIDSS quasars are shown as $\times$ symbols, with
redshift values encoded by symbol colour as in Figure \ref{colourcolourfig}.  A
quasar track, from $z=0.0$ (at top) to $z>3.5$ (at far right), with numeric
labels at intervals of $\Delta z =0.4$, derived from a modified version of the
quasar SED of \citet{maddox06} is also shown.}
\label{uzykfig} 
\end{figure} 

Here, we discuss the more tractable question of improving the effectiveness of
quasar selection at redshifts $z\sim2.5$ and $z\sim 5.6$ where colour-selected
samples in the optical encounter particular difficulties \citep{schneider05}.
The relatively deep UKIDSS, and future VISTA, limiting magnitudes in the
near-infrared, and the high rate of success in matching the present sample of
SDSS quasars to UKIDSS, suggests that quasars at $z\sim2.5$ and $z\sim 5.6$ may
be recovered by employing $K$-band information.  Figure \ref{jhkfig} shows, for
example, a more detailed view of the quasar-star separation in colour space when
$K$-band information is added.  While the population overlap (and thus
contamination) is significant for predicted quasar colours around $z\sim1.4$,
the colour separation becomes large from $z=2.0$ to $z=3.5$.  Because the
quasars shown are gathered from optical selection techniques, this implies that
independent (or joint) use of the near-infrared photometry on unidentified
stellar sources in such a region should yield additional quasars.

Thus, an appropriate choice of optical and near-infrared colours can select
quasars in the redshift ranges of interest.  Figure \ref{uzykfig} illustrates
one example of a successful quasar identification/separation scheme in the
$(u-z)$,$(Y-K)$ colour space which more cleanly separates normal Galactic stars
from quasars by sampling the underlying characteristic spectral break/inflexion
features, as described by \citet{warren00} and \citet{richards06b}.

This colour selection boundary illustrated is given by the following criteria: 
\begin{eqnarray}
& (u-z) < 4.5 &\\
& (Y-K) > 0.6 &\\
& (Y-K) > 0.35 ^* (u-z)+0.425,&
\end{eqnarray}

which results in a completeness of 97\,per cent (measured by the fraction of
quasars selected by the above cut) {\it for those quasars with magnitudes in
these bands}.  The overall success of the selection will vary as a function of
the magnitudes of the target population compared to the limiting magnitudes in
each passband.  For instance, the selection of $z>3$ quasars would be most
sensitive (and efficient) using colour cuts in the SDSS $g$ band and beyond,
such as ($r-Y$), ($Y-K$).

\begin{figure} \resizebox{0.48\textwidth}{!}{\includegraphics{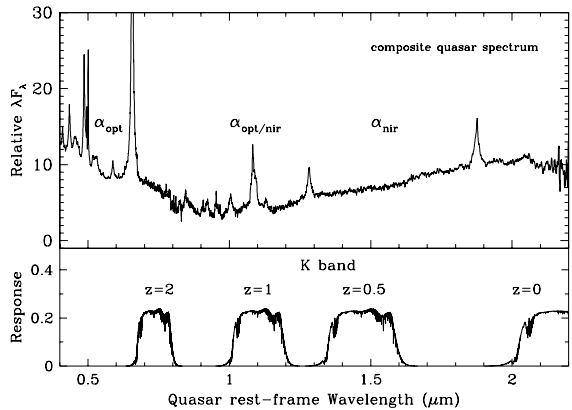}}
\vspace*{-2.2in} 
\caption{ As quasars of increasing redshift are selected using the
KX-technique, the $K$ filter probes increasingly blue rest-frame wavelengths of
the quasar SEDs.  As a result, different continuum slope regions
($\alpha_{\rm opt/nir}$) are sampled and it is important that appropriate
$K$-corrections are incorporated in any analysis of the properties of the
target population. A rest-frame quasar composite spectrum derived from
\citet{glikman06} and \citet{vanden01} is shown to illustrate the importance
of incorporating information about the SEDs of the target population(s).} 
\label{glikman}
\end{figure}

As illustrated above, the availability of photometry covering the rest-frame
optical and near-infrared wavelength regions offers the prospect of obtaining a
more complete census of the quasar population, both in terms of the redshift
range accessible and the variety of quasar SEDs included.  A consequence of the
extended restframe wavelength range accessed through obervations spanning the
$u$ through $K$ bands for quasars with redshifts $0 < z \la 7$, is the
importance of incorporating realistic model quasar SEDs in order to interpret
the properties of the quasar population reliably.  The simple power-law
approximation to quasar SEDs, modulated only by the presence of the strongest
emission lines, that is relatively effective when interpreting observations
probing only the restframe wavelength region $\lambda\lambda 1200-5000\,$\AA \
of high-luminosity quasars over redshifts $0.5 \la z \la 2.5$ is entirely
indequate when analysing quasar samples of the type presented here.  To
illustrate the point graphically, Figure \ref{glikman} shows a composite quasar
spectrum constructed from the ultraviolet composite of \citet{vanden01} and the
optical and near-infrared portion of the composite of \citet{glikman06}, joined
at 3100\,\AA.  The very large colour changes in the optical-near-infrared
colours as a function of redshift evident in Figure \ref{coloursvszfig} are a
direct manifestation of the form of the (very non-power-law-like) quasar SED.
Further care is required when considering the effect of host galaxy
contamination on the observed properties of the `quasars', particularly at
redshifts $z \la 1$ and at lower luminosities \citep{maddox06}.  The availability of the deep
near-infrared photometry would also allow a quasar sample to be selected over a
broader redshift range using observed passbands that closely approximate a fixed
restframe wavelength range for the quasars at each redshift, thereby minimising
the k-correction factors.  However, the more ambitious goal is now to utilise
the availability of the full set of $ugrizYJHK$ bands to define samples that
will provide powerful new constraints on the properties of the quasar population
as a function of redshift, luminosity and SED-type.

\section{Conclusion}

We have matched 98\,per cent of SDSS DR3 quasars over $\sim189$\,deg$^2$ with
their near-infrared counterparts in the UKIDSS DR1 source catalogue.  The
resulting sample of 2837 quasars is representative of the full range of
magnitudes and redshifts in the SDSS spectroscopic sample.  The astrometric
properties of the sample are similar to previous experience with 2MASS-SDSS
matched sources, while the UKIDSS photometry extends the matched imaging depth
by three magnitudes in the $JHK$ bands compared to 2MASS.  The median colours of
the matched quasars agree well with predictions generated using a model quasar
SED and provide a guide for higher redshift selection techniques.  
Optical and near-infrared colours, in appropriate combinations, provide
effective separation of stars and known quasars over essentially the full
redshift range $0 < z < 7$, overcoming the limitations of employing only optical
colours in the redshift ranges $z\sim2.5$ and $z\sim 5.6$.

\section{Acknowledgments} The authors are grateful for the efforts of all the UKIDSS 
consortium members who observed, processed, and produced the datasets analyzed
here.  We thank Steve Warren for
valuable information concerning the WFCAM imaging system, and the reviewer whose
comments improved the content of this work.  KC acknowledges
support from a UK PPARC rolling grant, and GTR has been supported in part by a
Gordon and Betty Moore Fellowship in Data Intensive Science at Johns Hopkins
University.  NM wishes to thank Corpus Christi College for the award of a Dr
John Taylor Scholarship.  UKIDSS observations are conducted from the United
Kingdom Infrared Telescope (UKIRT), which is operated by the Joint Astronomy
Centre on behalf of the U.K.  Particle Physics and Astronomy Research Council.
Funding for the creation and distribution of the SDSS Archive has been provided
by the Alfred P.  Sloan Foundation, the Participating Institutions, the National
Aeronautics and Space Administration, the National Science Foundation, the U.S.
Department of Energy, the Japanese Monbukagakusho, and the Max Planck Society.
The SDSS is managed by the Astrophysical Research Consortium (ARC) for its
Participating Institutions.

\bsp


\begin{thebibliography}{}

 
 
 
\bibitem[Abazajian et al.(2004)]{abz} Abazajian, K., et al. 2004, AJ, 128, 502
\bibitem[Allen et al.(2005)]{allen05} Allen, P.~R., Koerner, D.~W., Reid, I.~N., \& Trilling, D.~E.\ 2005, ApJ, 625, 385 
\bibitem[Brotherton et al.(2001)]{brotherton01} Brotherton M.~S., Tran H.~D., Becker 
R.~H., Gregg M.~D., Laurent-Muehleisen S.~A., White R.~L., 2001, ApJ, 546, 775 
\bibitem[Casali et al.(2001)]{casali01} Casali, M., \& et al.\ 2001, ASP Conf.~Ser.~232: The New Era of Wide Field Astronomy, 232, 357 
\bibitem[Chiu et al.(2005)]{chiu05} Chiu, K., et al.\ 2005, AJ, 130, 13 
\bibitem[Chiu et al.(2006)]{chiu06} Chiu, K., Fan, X., Leggett, S.~K., Golimowski, D.~A., Zheng, W., Geballe, T.~R., Schneider, D.~P., \& Brinkmann, J.\ 2006, AJ, 131, 2722 
\bibitem[Dye et al.(2006)]{dye06} Dye, S., et al. 2006, astro-ph/0603608
\bibitem[Emerson et al.(2004)]{emerson04} Emerson, J.P., et al. 2004, ESO Messenger No. 117  
\bibitem[Fan et al.(2006)]{fan06} Fan, X., et al.\ 2006, AJ, 
131, 1203 
\bibitem[Finlator et al.(2000)]{finlator00} Finlator, K., et al.\ 2000, AJ, 120, 2615
\bibitem[Francis et al.(1991)]{francis01} Francis P.~J., Hewett P.~C., Foltz C.~B., 
Chaffee F.~H., Weymann R.~J., Morris S.~L., 1991, ApJ, 373, 465 
\bibitem[Fukugita et al.(1996)]{fukugita} Fukugita, M., Ichikawa, T., Gunn, J.E., Doi, M., Shimasaku, K., \& Schneider, D.P. 1996, AJ, 111, 1748
\bibitem[Glikman et al.(2006)]{glikman06} Glikman, E., Helfand, D.~J., \& White, R.~L.\ 2006, ApJ, 640, 579 
\bibitem[Gunn et al.(1998)]{gunn98} Gunn, J.E., et al. 1998, AJ, 116, 3040
\bibitem[Hatziminaoglou et al.(2005)]{hatz05} Hatziminaoglou, E., et al.\ 2005, AJ, 129, 1198 
\bibitem[Hewett et al.(2006)]{hewett06} Hewett, P.~C., Warren, S.~J., Leggett, S.~K., \& Hodgkin, S.~T.\ 2006, MNRAS, 367, 454 
\bibitem[Hogg et al.(2001)]{hogg} Hogg, D.W., Finkbeiner, D.P., Schlegel, D.J., \& Gunn, J.E. 2001, AJ, 122, 2129
\bibitem[Lawrence et al.(2006)]{lawrence06} Lawrence, A., et al. 2006, astro-ph/0604426
\bibitem[Lupton et al.(1999)]{lupton} Lupton, R.~H., Gunn, J.~E., \& Szalay, A.~S. 1999, AJ, 118, 1406
\bibitem[Maddox \& Hewett(2006)]{maddox06} Maddox, N., \& 
Hewett, P.~C.\ 2006, MNRAS, 367, 717 
\bibitem[Pier et al.(2003)]{pier} Pier, J.R., Munn, J.A., Hindsley, R.B., Hennessy, G.S., Kent, S.M., Lupton, R.H., \& Ivezic, Z. 2003, AJ, 125, 1559 
\bibitem[Richards et al.(2002)]{richards02} Richards, G.T., et al. 2002, AJ, 123, 2945
\bibitem[Richards et al.(2006a)]{richards06a} Richards, G.~T., et al.\ 2006a, AJ, 131, 2766 
\bibitem[Richards et al.(2006b)]{richards06b} Richards, G.~T., et al.\ 2006b, astro-ph/0601558
\bibitem[Saunders et al.(2004)]{saunders04} Saunders, W., et al. \ 2004, Proc. SPIE, 5492, 410-420
\bibitem[Schneider et al.(2002)]{schneider02} Schneider, D.~P., et al. 2002, AJ, 123, 458
\bibitem[Schneider et al.(2005)]{schneider05} Schneider, D.~P., et al.\ 2005, AJ, 130, 367 
\bibitem[Skrutskie et al.(1997)]{skrutskie97}Skrutskie, M.~F., et al.\ 1997, in The Impact of Large Scale Near-IR Sky Surveys, ed.\ F.~Garz\'on, N.~Epchtein, A.~Omont, B.~Burton, and P.~Persei (Dordrecht: Kluwer), 25
\bibitem[Smith et al.(2002)]{smith} Smith, J., et al. 2002, AJ, 123, 2121 
\bibitem[Stern et al.(2005)]{stern05} Stern, D., et al.\ 2005, ApJ, 631, 163 
\bibitem[Stoughton et al.(2002)]{stoughton} Stoughton, C., et al. 2002, AJ, 123, 485
\bibitem[Strauss et al.(1999)]{strauss99}Strauss, M.~A., et al. 1999, ApJ, 522, L61
\bibitem[Vanden Berk et al.(2001)]{vanden01} Vanden Berk, D.~E., et al.\ 2001, AJ, 122, 549 
\bibitem[Vanden Berk et al.(2004)]{vanden04} Vanden Berk, D.~E.,  et al.\ 2004, ApJ, 601, 692 
\bibitem[Warren et al.(2006)]{warren06} Warren, S., et al. \ 2006, astro-ph/0610191
\bibitem[Warren \& Hewett(2002)]{warren02} Warren, S., \& Hewett, P.\ 2002, ASP Conf.~Ser.~283: A New Era in Cosmology, 283, 369 
\bibitem[Warren et al.(2000)]{warren00} Warren, S.~J., Hewett, P.~C., \& Foltz, C.~B.\ 2000, MNRAS, 312, 827 
\bibitem[White et al.(2006)]{white06} White, R.~L., et al. 2006, astro-ph/0607335
\bibitem[York et al.(2000)]{york} York, D. G., et al. 2000, AJ, 120, 1579
\bibitem[Zheng et al.(2000)]{zheng00} Zheng, W., et al. 2000, AJ, 120, 1607


\end{thebibliography}
\end{document}